\documentclass[aip,apl,reprint]{revtex4-1}
\usepackage{graphicx}
\usepackage{amsmath}
\usepackage{amssymb}
\usepackage{dcolumn}
\usepackage{bm}
\usepackage{amsfonts}
\usepackage{natbib}
\usepackage{hyperref}
\usepackage[papersize={8.5in,11in}]{geometry}

\begin{document}

\title{Quantum crossover in moderately damped epitaxial NbN/MgO/NbN junctions with
low critical current density}

\author{Luigi Longobardi}
\email{llongobardi@ms.cc.sunysb.edu}
\affiliation{Seconda Universit\'{a} degli Studi di Napoli, Dipartimento di Ingegneria dell'Informazione, via Roma 29, 81031 Aversa (Ce) Italy}
\affiliation{CNR-SPIN UOS Napoli, Complesso Universitario di Monte Sant'Angelo via Cinthia, 80126 Napoli (Na) Italy}
\author{Davide Massarotti}
\affiliation{Universit\'{a} degli Studi di Napoli Federico II, Dipartimento di Scienze Fisiche, via Cinthia, 80126 Napoli (Na) Italy}
\affiliation{CNR-SPIN UOS Napoli, Complesso Universitario di Monte Sant'Angelo via Cinthia, 80126 Napoli (Na) Italy}
\author{Giacomo Rotoli}
\affiliation{Seconda Universit\'{a} degli Studi di Napoli, Dipartimento di Ingegneria dell'Informazione, via Roma 29, 81031 Aversa (Ce) Italy}
\author{Daniela Stornaiuolo}
\affiliation{CNR-SPIN UOS Napoli, Complesso Universitario di Monte Sant'Angelo via Cinthia, 80126 Napoli (Na) Italy}
\author{Gianpaolo Papari}
\affiliation{NEST, CNR-NANO, and Scuola Normale Superiore di Pisa , Piazza dei Cavalieri  7, 56126 Pisa, Italy}
\author{Akira Kawakami}
\affiliation{Advanced ICT Research Institute, National Institute of Information and
Communications Technology, 588-2 Iwaoka-cho, Nishi-ku, Kobe 651-2492, Japan}
\author{Giovanni Piero Pepe}
\affiliation{Universit\'{a} degli Studi di Napoli Federico II, Dipartimento di Scienze Fisiche, via Cinthia, 80126 Napoli (Na) Italy}
\affiliation{CNR-SPIN UOS Napoli, Complesso Universitario di Monte Sant'Angelo via Cinthia, 80126 Napoli (Na) Italy}
\author{Antonio Barone}
\affiliation{Universit\'{a} degli Studi di Napoli Federico II, Dipartimento di Scienze Fisiche, via Cinthia, 80126 Napoli (Na) Italy}
\affiliation{CNR-SPIN UOS Napoli, Complesso Universitario di Monte Sant'Angelo via Cinthia, 80126 Napoli (Na) Italy}
\author{Francesco Tafuri}
\affiliation{Seconda Universit\'{a} degli Studi di Napoli, Dipartimento di Ingegneria dell'Informazione, via Roma 29, 81031 Aversa (Ce) Italy}
\affiliation{CNR-SPIN UOS Napoli, Complesso Universitario di Monte Sant'Angelo via Cinthia, 80126 Napoli (Na) Italy}

\date{\today}

\begin{abstract}
High quality epitaxial NbN/MgO/NbN Josephson junctions have been realized with MgO barriers up to a thickness of d=1 nm.  The junction properties coherently scale with the size of barrier, and low critical current densities  down to 3 A/cm$^2$ have been achieved for larger barriers. In this limit, junctions exhibit macroscopic quantum phenomena for temperatures lower than 90 mK. Measurements and junction parameters support the notion of a possible use of these devices for multiphoton quantum experiments, taking advantage of the fast non equilibrium electron-phonon relaxation times of NbN.
\end{abstract}


\maketitle

Advances in the quality of the traditional Josephson structures based on Nb and Al technology as well as the search for high quality junctions of different materials are a major contribution in superconducting electronics. In addition they may respond to a wider spectrum of qubits requirements and functionalities, and may contribute to investigate quantum regimes\cite{caldeira1983,clarke2008}. We inquire on the quantum behavior of low critical current density ($J_c$) NbN/MgO/NbN Josephson junctions (JJs). These kinds of JJs are usually aimed to high $J_c$ values of about $10kA/cm^2$ and used in superconducting electronic circuits \cite{Yamamori,Larrey}.
In this work we show that the NbN junction technology can be extended to low $J_c$ values of about $3 A/cm^2$, which are two to four orders of magnitude lower than previously reported values  \cite{kawakami_2003}. NbN is a material of great interest for sensor applications\cite{ejrnaes09,marsili11,delacour,Verevkin} and it guarantees both fast non equilibrium electron-phonon relaxation times  ($\tau$) and higher gap values, when compared with traditional junction technologies based on Nb, Al and Pb \cite{kaplan_1976}.

\begin{figure}
\includegraphics[width=6.5cm]{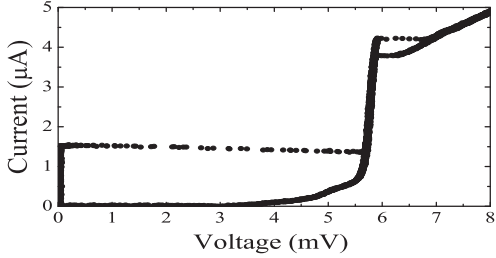}
\caption{Current-Voltage characteristic of a $10 \mu m$ diameter JJ measured at 290mK.}
\label{fig:fab}
\end{figure}

In this work, a canonical quantum behavior typical of a moderately damped junction is observed along with a crossover temperature  between the thermal and quantum regime of about 90 mK. We found a phase diffusion regime, with a damping parameter $Q\sim3$ consistent with what observed in various Josephson systems\cite{pekola2005}. The possibility to achieve very low critical current densities in moderately damped NbN junctions, combined with their short electron relaxation time  $\tau < 10 ps$  \cite{il'in_2000}, paves the way to experiments aimed at observing extreme multi-photon tunneling quantum effects \cite{Ivlev_2005} thus extending studies performed up to now on few photons processes\cite{ustinov2003,saito,bauch_2006}. The intrinsic properties of low-Jc NbN junctions with maximum critical current $I_{co} < 1 \mu A$ meet the condition  $\tau < \hbar/E_J =2e/I_{co}$, where $E_J=\hbar I_{co} / 2 e$ is the Josephson energy, nominally required to observe multi-photon effects in the presence of a non-stationary signal. These parameters would guarantee accessible working regimes to possibly observe Euclidean resonance (ER) \cite{Ivlev_2005}. According to numerical simulations presented by Ivlev et al.\cite{Ivlev_2005} for ER proposals, our moderately damped junction would require amplitudes  ($\tilde{I}$)  and frequencies ($\Omega_p$) of the non stationary component of the bias current lower than 0.2 $\mu A$ and larger than 20 GHz respectively, which are experimentally feasible values.

The JJs used in this work consist of an epitaxially grown NbN/MgO/NbN trilayer\cite{kawakami_jap}. Epitaxial MgO is traditionally a high quality tunnel barrier \cite{kline} whose complete impact on superconducting electronics is still under investigation and depends on the interface with the superconducting layers, on its epitaxial quality and possibly on its thickness.

In the present work, the NbN base (BE) and counter electrode (CE) are both 200nm thick and were deposited on a single-crystal MgO substrate at ambient substrate temperature\cite{zang1996} using DC magnetron sputtering with a Nb target in a mixture of 5 parts argon and 1 part nitrogen gases. The MgO tunnel barrier is about 1.0nm thick and was deposited by rf sputtering  using a low incident rf-power density in order to avoid damaging of the surface of the base electrode. This step was followed by junction definition using a reactive ion etch (RIE) followed by the deposition of a MgO insulating layer patterned by a lift-off process. Finally, a wiring layer was realized by deposition of 350nm of NbN which was then patterned and defined by RIE.
The realized junctions have a superconducting transition temperature of about 16.6 K for both electrodes. A typical example of current-voltage characteristics is shown in fig. \ref{fig:fab} at the temperature of 290 mK. We measured circular junctions of various diameters ranging from 2 to 10$\mu m$ and consistently found a critical current density of $3 A/cm^2$. Large values of the gap voltage ($V_g=5.7mV$) and of the $I_{co}R_N$ product, with $R_N$ the normal state resistance,  of about 5mV have been measured along with a small subgap leakage voltage ($V_m=23.5mV$ measured at 3mV)\cite{shoji_1098}. From the magnetic field dependence of the critical current we estimated a London penetration depth ($\lambda_L$) of about 190nm meanwhile the Josephson  penetration depth was $\lambda_J = 150 \mu m$. This nominally guarantees more uniform currents than in devices of comparable sizes but larger $J_c$. It should be noted that many measurements of $\lambda_L$ using resonant methods have been reported in literature\cite{moodera1985,oates1991,komiyama1996}. Our value is nearly a factor of 2 lower than what found for near-stoichiometric polycrystalline films \cite{moodera1985,oates1991} but it is in very good agreement with the value obtained on a single crystal NbN film\cite{komiyama1996}.

The dynamics of a current-biased Josephson junction is governed by the phase difference $\varphi$ of the order parameter across the junction, as extensively described in literature\cite{barone,devoret1985} and is equivalent to the classical motion of a particle in a washboard potential $U(\varphi)=-E_J(\cos\varphi +\frac{I}{I_{co}}\varphi)$ with damping $Q^{-1}=(\omega_{po} R C)^{-1}$, where R and C are the junction resistance and capacitance, and $\omega_{po}=(2eI_{co}/\hbar C)^{1/2}$ is the zero bias plasma frequency.  For $k_b T \gg \hbar \omega_{po}$ the escape process is dominated by thermal activation with a rate $\Gamma_T= a_t (\omega_p/2\pi)\exp \left(-\Delta U /k_B T\right)$ where $a_t=4/[(1+Q k_b T / 1.8 \Delta U)^{1/2}+1]^2$ is the thermal prefactor. At low enough temperature the escape is dominated by Macroscopic Quantum Tunneling (MQT) \cite{devoret1985} with a rate $\Gamma_q= a_q (\omega_p/2\pi)\exp \left[-7.2 \Delta U/\hbar \omega_p\left(1+ 0.87/Q\right)\right]$ where $a_q =[120\pi(7.2\Delta U/\hbar \omega_p)]^{1/2}$.
\begin{figure}
\includegraphics[width=6.5cm]{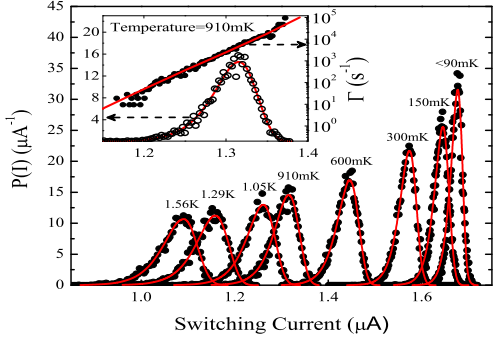}
\caption{Switching current probability distribution at $H=0G$ for different bath temperatures. The symbols represent data, and the lines are fit. The inset shows the quality of the fit at T=910mK, along with the data converted into escape rates.} \label{fig:fit}
\end{figure}
To study the escape rates of a NbN/MgO/NbN JJ we have measured the switching current probability of a JJ with 10$\mu m$ diameter in a dilution refrigerator with a loaded base temperature of 20mK.
The current bias source was coupled through battery powered unity gain isolation amplifiers in order to be effectively disconnected from the earth ground\cite{bennett2009}. Filtering was provided by low pass filters, with a cut-off frequency of 1.6MHz thermally anchored at 1.4 K, and by a combination of copper powder \cite{milliken:024701} and twisted pair filters \cite{spietz} thermally anchored at the mixing chamber. A room temperature electromagnetic interference filter stage has been also used for the junction current and voltage lines. The junction was biased with a sweep rate $dI/dt=122 \mu A/s$ and at least $10^4$ switching events have been recorded using a standard technique \cite{doug2007}.

The moderately damped nature of these junctions generates a characteristic diffusive phase dynamics\cite{pekola2005}. Here we focus on the MQT regime. In figure \ref{fig:fit} we show the normalized switching current distribution at different temperatures along with a theoretical fit. For these fits we obtained $I_{co}=1.91 \pm 0.03  \mu A$, while the value of the damping, $Q=2.7 \pm 0.1 $, is deduced by fitting the escape rates at higher temperatures, in the phase diffusion regime, using a Monte Carlo simulation of the escape and retrapping process\cite{pekola2005}.

\begin{figure}
\includegraphics[width=6.5cm]{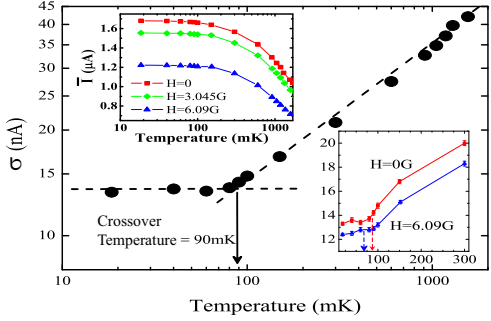}
\caption{Temperature dependence of the standard deviation, $\sigma$, of the switching distributions at zero magnetic field; the error bars are within the width of the data point. The top inset displays the switching distributions mean current vs temperature for three values of applied magnetic field. The bottom inset shows $\sigma$ vs temperature below 300mK for two values of magnetic field. The arrows indicate the crossover temperature between the thermal and quantum regimes.} \label{fig:sigma}
\end{figure}

In fig. \ref{fig:sigma} we report the width of the distribution, $\sigma$, vs temperature. Above 90mK the data agree with predictions for thermal activation, the saturation of $\sigma$ at $T<90mK$ indicates that the escape is dominated by quantum tunneling.  The temperature dependence of $\sigma$ for a reduced critical current when an external magnetic field  H is applied to the junction, is a neat confirmation of MQT (bottom right inset) \cite{devoret1985}. H induces a decrease of $\omega_P$ and therefore of the crossover temperature   $T_{cross}=\frac{\hbar \omega_{po}}{2\pi k_B} ((1+\frac{1}{4Q^2})^{1/2}-\frac{1}{2Q})$ consistently with MQT expectations\cite{devoret1985,barone07}.
An estimate of the capacitance $C=0.3 pF$ and of the plasma frequency $\omega_{po} \simeq$ 22 GHz can be obtained on the basis of  $T_{cross}$ and $I_{co}$, and therefore of the measurements shown in figures \ref{fig:fit} and \ref{fig:sigma}. The value of the capacitance can in principle be calculated also from the position in voltage of the Fiske steps, but in our case, the amplitude of such steps is vanishingly small \cite{barone}. This is partly due to the fact that the Fiske resonance amplitude depends on\cite{nerenberg1974} $J_c*(r/\lambda_J)^2$, where r is the junction radius, therefore it fades out because of the low values of $J_c$ found in our junctions and because $r \ll \lambda_J$. The discrepancy between the value of the capacitance obtained from MQT measurements and what calculated from Fiske steps is a long standing debate\cite{devoret1985,barone}.

In conclusion MgO barriers of thickness of about 1.0 nm guarantee moderately damped NbN Josephson junctions with macroscopic quantum behavior for temperatures lower than 90 mK. The devices are characterized by  a set of parameters promising for the realization of systematic multiphoton quantum experiments, which take advantage of the fast non equilibrium electron-phonon relaxation times of NbN common also to high $T_c$ JJs\cite{Ivlev_2005,bauch_2006}. The moderately damped nature of the junctions, further favored by a decrease in its size guarantees an experimentally feasible set of junctions parameters, which can be further controlled during ER experiments.

We acknowledge the support of the European Commission through STREP "MIDAS-Macroscopic Interference Devices for atomic and Solid State Physics: Quantum Control of Supercurrents" and a Marie Curie International Reintegration Grant within the 7th Framework Programme.

\end{document}